\pdfoutput=1

\documentclass[reprint,aps,prl,showpacs,twocolumn]{revtex4-1}
\usepackage{mathptmx}
\usepackage{epsfig,amsopn}
\usepackage{graphicx}
\usepackage{amsmath,amssymb}
\usepackage{natbib,color}
\usepackage{braket}
\usepackage{float}
\usepackage{enumerate}
\allowdisplaybreaks[1]

\newcommand{\eref}[1]{Eq.~(\ref{#1})}%
\newcommand{\fref}[1]{Fig.~\ref{#1}} %

\def\bea{\begin{eqnarray}}
\def\eea{\end{eqnarray}}

\def\nn{\nonumber}

\def\f{\frac}

\begin{document}

\title{First passage under restart with branching}



\author{{\normalsize{}Arnab Pal$^{1,2,3}$}
{\normalsize{}}}
\email{richard86arnab@gmail.com}

\author{{\normalsize{}Iddo Eliazar}
{\normalsize{}}}
\email{eliazar@post.tau.ac.il}

\author{{\normalsize{}Shlomi Reuveni$^{1,2,3}$}
{\normalsize{}}}
\email{shlomire@tauex.tau.ac.il}

\affiliation{\noindent \textit{$^{1}$School of Chemistry, Raymond and Beverly Sackler Faculty of Exact Sciences, Tel Aviv University, Tel Aviv 6997801, Israel}}

\affiliation{\noindent \textit{$^{2}$Center for the Physics and Chemistry of Living Systems. Tel Aviv University, 6997801, Tel Aviv, Israel}}

\affiliation{\noindent \textit{$^{3}$The Sackler Center for Computational Molecular and Materials Science, Tel Aviv University, 6997801, Tel Aviv, Israel}}

\date{\today}

\begin{abstract}

\noindent{First passage under restart with branching is proposed as a generalization of first passage under restart. Strong motivation to study this generalization comes from the observation that restart with branching can expedite the completion of processes that cannot be expedited with  simple restart; yet a sharp and quantitative formulation of this statement is still lacking. We develop a comprehensive theory of first passage under restart with branching. This reveals that two widely applied measures of statistical dispersion---the coefficient of variation and the Gini index---come together to determine how restart with branching affects the mean completion time of an arbitrary stochastic process. The universality of this result is demonstrated and its connection to extreme value theory is also pointed out and explored.}

\end{abstract}

\maketitle

Restart can expedite the completion of a myriad of processes \cite{Restart1,Restart2,Restart3,Restart-Biophysics1,Restart-Biophysics2,Restart-Biophysics3,Restart4,Restart5,ReuveniPRL,PalReuveniPRL,Restart-Search1,Restart-Search2,Restart-Search3}. It works by taking advantage of stochastic fluctuations to trade long completion times with ones which are shorter on average. Thus, restart is particularly effective in expediting processes whose statistical completion time distributions are very broad. A classical example of this principle is the first passage time (FPT) \cite{FPT1,FPT2} to a stationary target of a particle undergoing diffusion \cite{RednerBook,MetzlerBook,Schehr-review}. The distribution of this FPT is so broad that its mean diverges \cite{RednerBook}. On average, a diffusive particle will thus take an infinite amount of time to reach a target; but restart can change this situation dramatically. Indeed, stopping the particle's motion intermittently and placing it back at the origin to continue its motion from there regularizes the FPT distribution and renders its mean finite \cite{Restart1,Restart2}. Remarkably, this `regularization by restart' works for any process with divergent mean completion times -- regardless of the specific details \cite{Restart-Biophysics1,Restart-Biophysics2,ReuveniPRL,PalReuveniPRL}. 

Some processes cannot be expedited with restart as this strategy crucially depends on the existence of stochastic fluctuations in completion times. Consider, as an illustrative example, a train which travels from point A to B at a constant velocity. Aside from small glitches here and there, such train takes a fixed amount of time to reach its final destination. Thus, the train's FPT distribution is very sharply peaked around its mode. As it turns out, this punctuality is intimately related with the fact that taking the train, at any point along its route, and placing it back at the dock of departure will only prolong its journey. Indeed, whenever the coefficient of variation ($CV$) of the completion time is smaller than one -- applying restart will increase the mean completion time \cite{Restart-Biophysics1,Restart-Biophysics2,ReuveniPRL,PalReuveniPRL}. As the $CV$ is the ratio of the standard deviation to the mean, this criterion captures the need for relatively large stochastic fluctuations; also, this criterion sets a sharp boundary for the application of restart as a `speed-up' mechanism. 

\begin{figure}
 \begin{centering}
\includegraphics[width=7.cm]{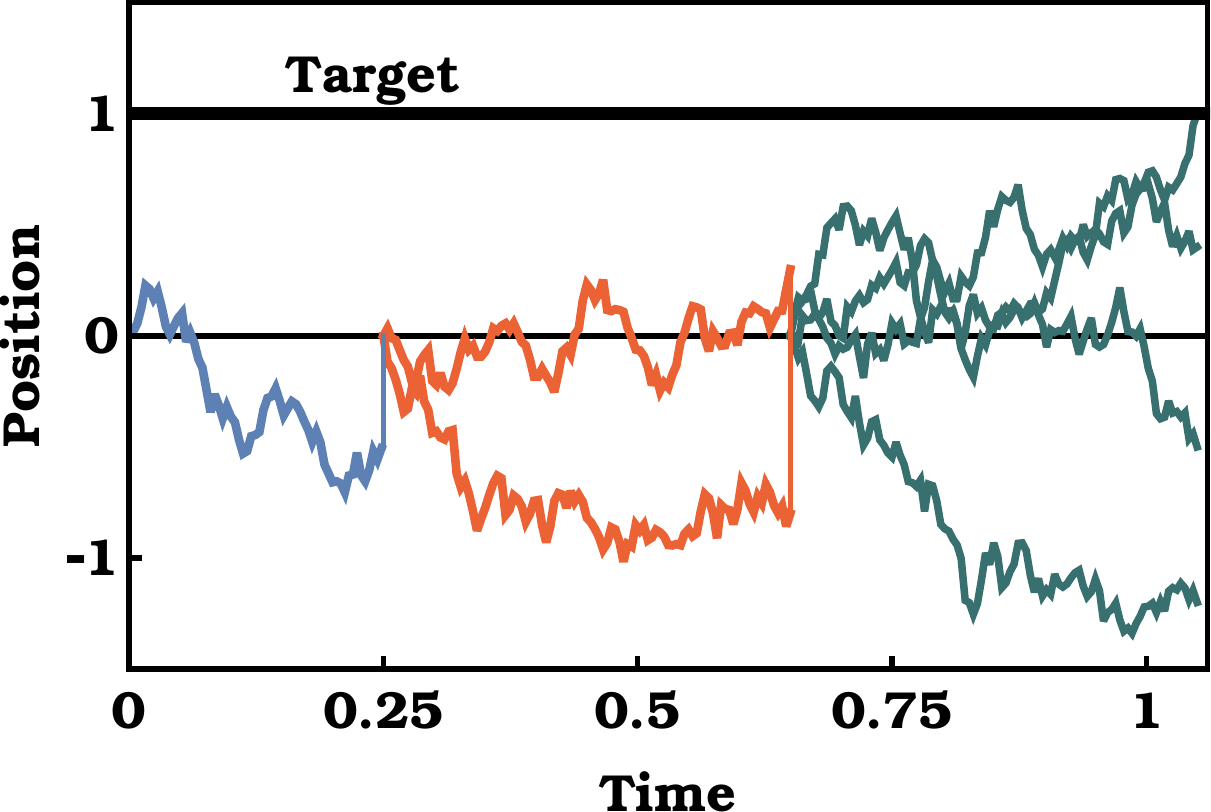}
\end{centering}
\caption{An illustration of diffusion under restart with branching. When restart occurs, particles are brought back to the origin and each particle is branched into two identical copies of itself.}
\label{fig1}
\end{figure}

Branching a process -- so as to create several competing copies of it that run parallel to one another -- is a different `speed-up' strategy that is used, e.g., in computer science \cite{parallel-processing1,parallel-processing2}. Branching allows one to reduce the time taken to solve a problem at the expense of resource investment. Restart is also used for that purpose \cite{restart-CS1,restart-CS2}, and while it is not always applicable it could generate speedup without requiring additional resources. A strategy that combines restart and branching may thus allow one to have the best of both worlds. 

Branching was extensively studied, e.g., in birth and death processes \cite{Harris-Book}, reaction-diffusion systems \cite{Bramson,Derrida}, Brownian motion \cite{Kabir-1,Kabir-2}, and epidemic spread \cite{Satya-PNAS}. However, it was just earlier this year that branching was considered in the context of restart---coining the term `branching search' \cite{Iddo-Branch}. To illustrate this, envision a scout that is dispatched by a central command to search for a hidden target, e.g., treasure, an enemy base, or a group of missing people. If the scout fails to find the target within a certain time window then central command calls the scout back, and dispatches two new scouts to the task. If the two also fail within a certain time window then they are also called back, and four new scouts are dispatched to the task. Thus, instead of merely resetting the search in an attempt to expedite it, efforts to locate the target are doubled with every failed attempt. 

A scenario of the type described above could be mathematically modeled by, e.g., considering diffusion under stochastic restart with branching (\fref{fig1}); yet this model does not fall under the existing theoretical framework. Moreover, while our understanding of first-passage under restart enjoys the aforementioned $CV$ criterion, a universal criterion that determines the effect of `restart-\&-branching' on the mean completion time of a general stochastic process is currently missing. Indeed, while the great advantages that restart brings to existing search strategies \cite{Shlesinger,Benichou-review} were already discussed extensively \cite{Restart1,Restart-Biophysics2,PalReuveniPRL,Restart-Search2,Restart-Search3,Search-DNA,Chechkin}, the exploration of 'branching search' \cite{Iddo-Branch} has just begun---determining the combined effect of restart and branching is a non-trivial task even for the relatively simple case illustrated in Fig. 1.

\textit{Diffusion under restart with branching.}---Consider a particle undergoing diffusion with drift. The particle starts its motion at the origin and drift-diffuses at an average velocity $V>0$ until it hits a target located at $L>0$. This will happen at a random time $T$ which comes from the Inverse-Gaussian distribution $f_T(t)=\frac{L}{\sqrt{4\pi D t^3}} e^{-\frac{(L-Vt)^2}{4Dt}} (t\geq0)$, with $D$ standing for the diffusion coefficient \cite{Cox-Miller-Book}. Defining the P\'eclet number $Pe=LV/2D$, one can then trivially observe that diffusion (drift) becomes negligible in the limit of $Pe\gg1$ ($Pe\ll1$). Indeed, $f_T(t)$ becomes sharply peaked for high P\'eclet numbers and resembles that of a particle undergoing pure diffusion when P\'eclet numbers are low (\fref{figS1}). Thus, by tuning the P\'eclet number, one can explore the wide range of cases that stand between the two extreme scenarios discussed above: simple motion at a constant velocity and pure diffusion.    

Let us now consider the above process, but with the addition of restart and branching (\fref{fig1}). Specifically, assume that restart can occur with some probability at any given point in time; and that when this happens our particle is taken back to the origin and branched into two, or generally $m$, identical particles. These particles will start their motion anew, but if restart occurs again they will all be taken back to the origin and branched as before. This procedure repeats itself until one of the particles in the system hits the target.

\begin{figure}
\centering
\includegraphics[width=7cm]{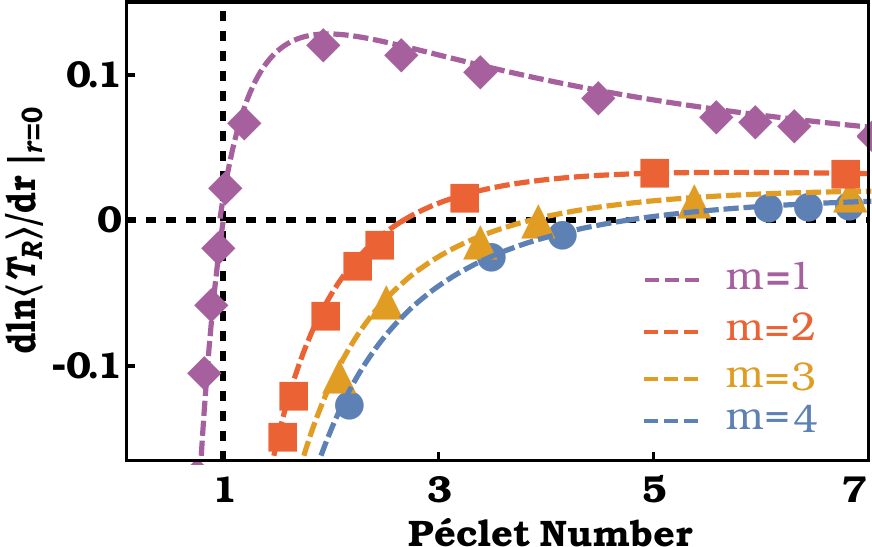}
\caption{The effect the introduction of restart with branching has on the mean FPT (to a stationary target) of a particle undergoing diffusion with drift. Results are shown for different values of the P\'eclet number, $Pe=LV/2D$, and branching index $m$.}
\label{fig2}
\end{figure}

To study the effect of restart with branching, we simulated the process illustrated in \fref{fig1} \cite{SM}. First, we wanted to explore what happens when restart with branching is introduced into the system. Letting $r$ denote the constant rate at which restart with branching occurs, we looked at the logarithmic derivative of the mean FPT to the target with respect to $r$ at $r=0$. This was plotted vs. the P\'eclet number for different values of the branching parameter $m$ (\fref{fig2}, markers). A negative derivative indicates that the introduction of restart with branching will, on average, expedite first passage to the target, and the converse is true when a positive derivative is found. Results coming from simulations were later on corroborated by the general theory developed below (\fref{fig2}, dashed lines).

Analyzing the results presented in \fref{fig2}, we would first like to draw attention to the case $m=1$. In this degenerate case restart is not accompanied by branching and one can observe that the derivative changes sign exactly when $Pe=1$. This is no surprise. The mean of the Inverse-Gaussian distribution above is $L/V$ and its standard deviation is $\sqrt{2DL/V^3}$. Letting $CV$ stand for the coefficient of variation of this distribution, we observe that $CV^2=1/Pe$. Recalling that the theory of first passage under restart asserts that restart will expedite the completion of a process if and only if $CV>1$ \cite{Restart-Biophysics1,Restart-Biophysics2,ReuveniPRL,PalReuveniPRL}, we conclude that the observed behaviour is a special case of a more general one. 

For $m>1$, we see that the critical P\'eclet number at which the derivative changes sign is always larger than unity. Moreover, this critical number grows as $m$ increases. Thus, in the case of diffusion with drift, branching allows restart to reduce the mean FPT even when the underlying FPT distribution is much narrower than the critical limit for simple restart ($Pe=1$). However, and in contrast to the $m=1$ case, we find the corresponding set of critical $CV$s (\fref{figS2}) to be particular to diffusion with drift, i.e., non-universal. Any general characterization of the effect that restart with branching has on the mean completion time of a generic process must thus rely on more than the first two moments of the underlying time distribution. To find this characterization, and further deepen our understanding, we now develop a general theory of first passage under restart with branching.     

\begin{figure}[t]
\includegraphics[width=8cm]{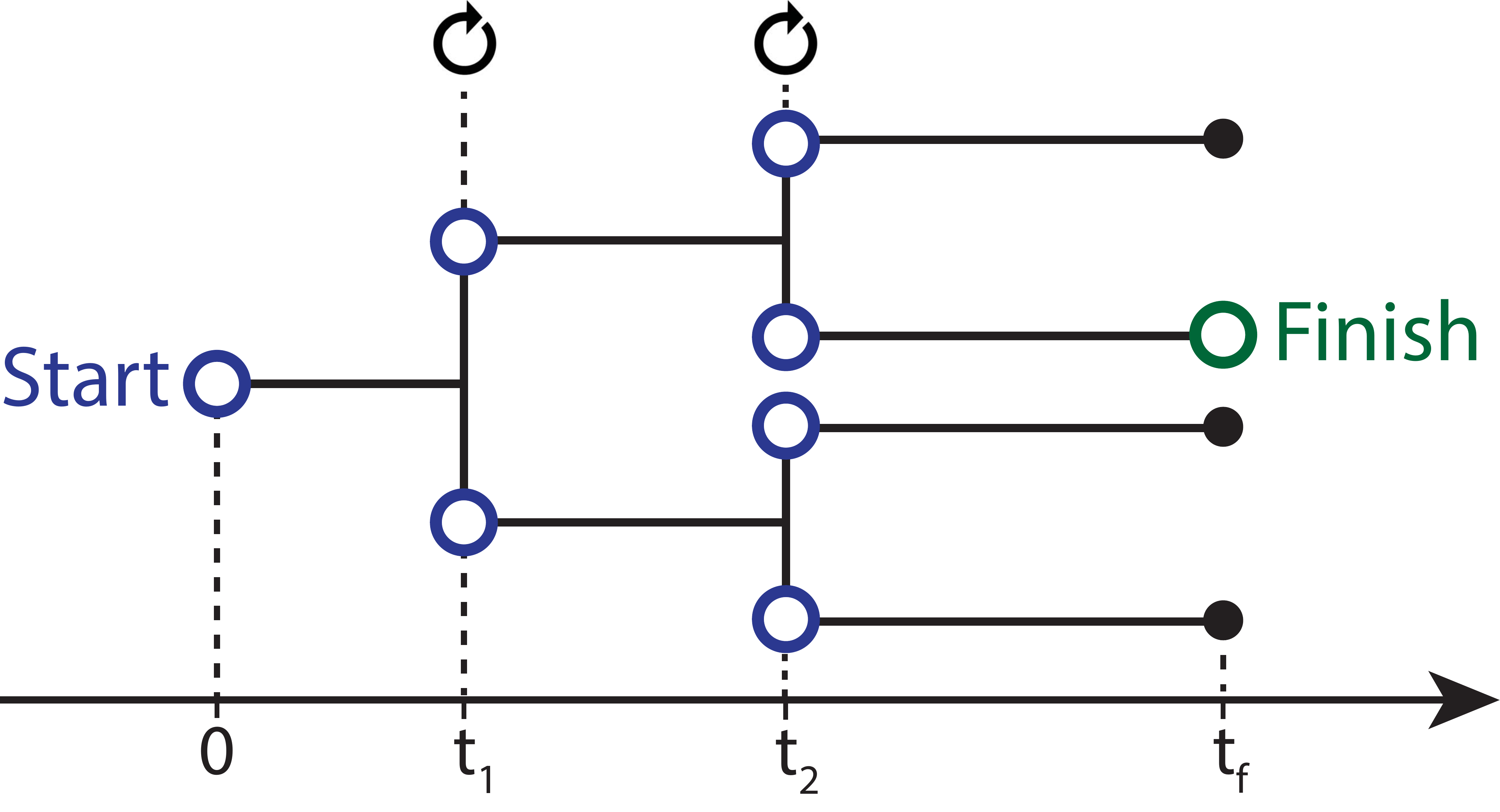}
\caption{An illustration of first passage under restart with branching ($m=2$). Restart and branching takes place at time $t_1$ and thereafter at time $t_2$. Completion takes place at time $t_f$.}
\label{fig3}
\end{figure}

\textit{A theory of first passage under restart with branching.}---Consider a generic process that starts at time zero and, if allowed to take place without interruptions, completes after a random time $T$. The process, however, can also be restarted after some random time $R$ and consecutively branched into $m$ daughter processes which are independent and identical copies of the parent process  $(m=1,2,3,\cdots)$. Thus, if restart prevents the parent process from completing, $m$ daughter processes will start in its stead. The same procedure will then repeat itself: if one of the daughter processes is able to finish prior to restart completion will be declared. Otherwise, following a second restart event, each daughter process will, in itself, be branched into $m$ processes that will once again start their course afresh; and so on and so forth until one of the offspring processes is able to complete (\fref{fig3}). 

To analyze the above scenario, we let $R'$ and $\{T_{1},\cdots,T_{m}\}$ denote independent and identically distributed copies of $R$ and $T$ respectively. We then observe that $T_{R}$---the random completion time of a generic process under restart with branching---can be written as \noindent 
\begin{equation}
\begin{array}{l}
T_{R}=\left\{ \begin{array}{lll}
T &  & \text{if }T<R\text{ }\\
 & \text{ \ \ }\\
R+[T^{(m)}]_{R'} &  & \text{if }R\leq T\text{ ,}
\end{array}\right.\text{ }\end{array}\label{renewal-1}
\end{equation}
with $T^{(m)}=\text{min}\{T_1,\cdots,T_m\}$.
The rational leading to \eref{renewal-1} is simple. If the parent process is able to complete prior to restart the process there ends and $T_{R}=T$. Otherwise, after some random time $R$, restart with branching occurs. One could then `virtually' join all the newly formed daughter processes into a single process whose uninterrupted completion time is $T^{(m)}$, i.e., the minimum over $m$ IID copies of $T$. However, since this joint process is also subject to restart with branching its actual completion time is given by $[T^{(m)}]_{R'}$. Thus, when restart precedes completion, we have $T_{R}=R+[T^{(m)}]_{R'}$. 

\eref{renewal-1} can be used to derive a formula for the mean FPT of a process that has been subjected to restart with branching. This is done in iterative steps. First, observe that \eref{renewal-1} implies that  $T_R=\text{min}(T, R)+I(R \leq T)[T^{(m)}]_{R'}$, 
where $\text{min}(T, R)$ is the
minimum of $T$ and $R$, and $I(R \leq T)$ is an indicator random
variable which takes the value one when $R \leq T$ and zero otherwise. Taking expectations, and utilizing statistical independence, we obtain that
\bea
\langle T_R \rangle=\langle \text{min}(T, R) \rangle+\text{Pr}(R \leq T) ~\langle [T^{(m)}]_{R} \rangle \text{ .}
\label{renewal-2}
\eea
The first term in \eref{renewal-2} and the probability that $R \leq T$ can then be evaluated given the distribution functions of $T$ and $R$ \cite{SM}. To further proceed, we utilize \eref{renewal-1} one more time to obtain
$[T^{(m)}]_{R}=\text{min}(T^{(m)}, R)+I(R \leq T^{(m)})[T^{(m^2)}]_{R'}$ \cite{SM}. 
Taking expectations one arrives at an equation similar to \eref{renewal-2} \cite{SM} where $\text{min}(T^{(m)}, R)$ and the probability that $R \leq T^{(m)}$ can once again be directly computed given the distribution functions of $T$ and $R$. As for the expectation of $[T^{(m^2)}]_{R'}$, this is again evaluated by iterative use of \eref{renewal-2}. 

\begin{figure}[t]
\includegraphics[width=6.5cm]{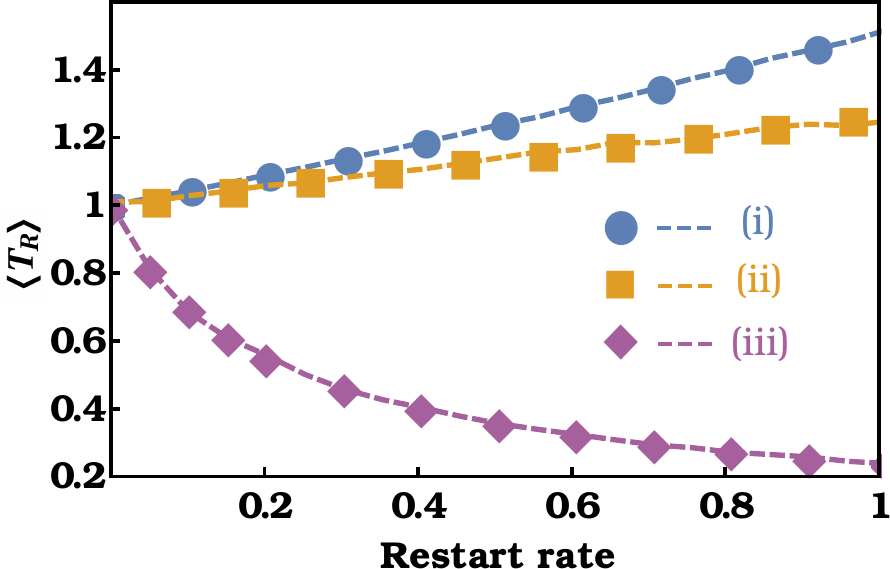}
\caption{Mean FPT under restart with branching vs. restart rate for different FPT processes: (i) cover time of a random walker \cite{Benichou-covertime}; (ii) diffusion with drift \cite{Cox-Miller-Book}; and (iii) an infectious cells growth model \cite{Iyer-Biswas}. In all three cases, the branching index was set to $m=2$, and time was measured in units of the mean FPT in the absence of restart and branching. Dashed lines are theoretical curves computed based on \eref{MFPTUB} and markers come from Monte Carlo simulations (\cite{SM}).}
\label{fig4}
\end{figure}

Carried ad infinitum, the above procedure yields a formula that allows the evaluation of $\langle T_R \rangle$ for arbitrary choices of $T$ and $R$ \cite{SM}. Of particular interest in that regard is the case where restart is conducted at a constant rate $r$, i.e., when the random restart time $R$ is exponentially distributed with mean $1/r$ \cite{Restart1,Restart2,Restart-Biophysics1,Restart-Biophysics2,ReuveniPRL}. We then find 
\bea
\langle T_R \rangle=\frac{1}{r}~\sum_{n=0}^{\infty}~\prod_{k=0}^{n}~[1-\tilde{T}^{(m^k)}(r)]~,
\label{MFPTUB}
\eea
with $\tilde{T}^{(m^k)}(r)$ standing for the Laplace transform of $T^{(m^k)}$ evaluated at $r$ \cite{SM}. Equation (\ref{MFPTUB}) allows one to study the effect of restart with branching on the mean completion times of various process by systematic evaluation of $\langle T_R \rangle$ (\fref{fig4}). Depending on the process, and the branching index $m$, the introduction of restart with branching could then either increase or decrease the mean completion time. A universal and clear cut criterion to determine which of the two occurs is therefore in need. We will now derive this criterion and show that aside from the coefficient of variation it is also sensitive to another measure of statistical dispersion: the generalized Gini index.

The Gini index is widely applied in economics and in the social sciences as a measure of socioeconomic inequality \cite{Gini-methodology}. In general, the Gini index is a measure of statistical dispersion that is applicable to  non-negative valued random variables with positive means \cite{Iddo-Tour}. Here we apply the Gini index in the context of the random variable $T$. The Gini index has several equivalent representations, and one of them is \cite{Iddo-Tour} $GI=1-\langle \min \left\{ T_{1},T_{2}\right\}\rangle/\langle T \rangle$, where $T_1,T_2$ are two IID copies of $T$. In turn, this representation can be generalized from two IID copies of $T$ to $m$ copies, $\{T_{1},\cdots
,T_{m}\}$, yielding the following generalized Gini index \cite{Iddo-Physica}
\begin{equation}
GI_{m}=1-\langle \min \left\{ T_{1},\cdots
,T_{m}\right\} \rangle/\langle T \rangle \text{ ,}  \label{Gini}
\end{equation}
$(m=1,2,...)$. Note that the index of order one vanishes $GI_{1}=0$; and that the index of order two is the Gini index $GI_{2}=GI$.

With the generalized Gini index at hand, and the coefficient of variation $CV=\sigma(T)/\langle T \rangle$, we find that $\f{d\langle T_R \rangle}{dr}|_{r=0} < 0$ if and only if \cite{SM} \bea
CV^2 + 2\cdot{GI}_{m} >~ 1 \text{ .}
\label{main}
\eea
Equation (\ref{main}) provides a universal criterion that determines how restart with branching affects the mean completion time of an arbitrary stochastic process. When $m=1$, i.e., when there is no branching, Eq. (\ref{main}) boils down to the known `pure restart' criterion $CV>1$ 
\cite{Restart-Biophysics1,Restart-Biophysics2,ReuveniPRL,PalReuveniPRL}. However, for $m>1$, we see that the criterion in \eref{main} also depends on the generalized Gini Index. To better illustrate this we now consider a concrete example.

\begin{figure}
\centering
\includegraphics[width=8.5cm]{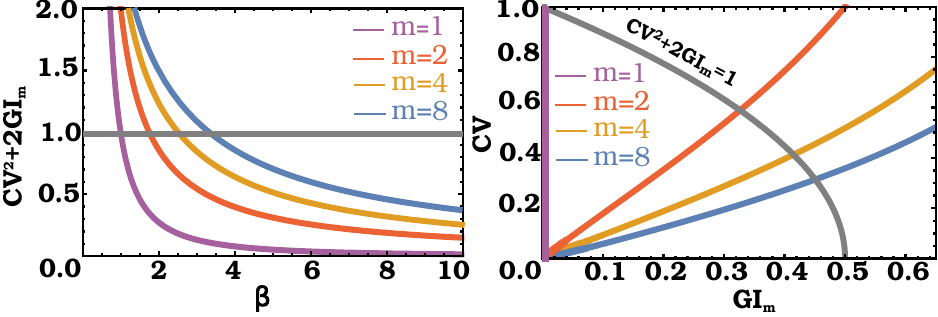}
\caption{Graphical exploration of the criterion from Eq. (\ref{main}) when applied to the Weibull distribution. Left. $CV^2 + 2{GI}_{m}$ is plotted vs. the shape parameter $\beta$ for different values of the branching index $m$. Right. The $CV$ is plotted vs. $GI_m$ for different values of the branching index $m$. In both panels, the  $CV^2+2{GI}_{m}=1$ line separates the two different behavioural phases that are predicted by Eq. (\ref{main}).}
\label{fig5}
\end{figure}

Consider the  Weibull distribution $\text{Pr}[T \geq t] = e^{-(t/\alpha)^{\beta}}~(t\geq 0)$; with $\alpha,\beta>0$ standing respectively for the scale and shape parameters. The coefficient of variation of this distribution is known to be given by $CV^2=\Gamma(1+2/\beta)/\Gamma(1+1/\beta)^{2}-1$, where $\Gamma(x)$ is the Gamma function. To compute $GI_m$, we first observe that if $T$ is a Weibull random variable with parameters $\alpha$ and $\beta$, then $T^{(m)}=\min \left\{ T_{1},\cdots,T_{m}\right\}$ is also a Weibull random variable with scale parameter $\alpha/m^{1/\beta}$ and shape parameter $\beta$ \cite{SM}. \eref{Gini} then gives ${GI}_{m}=1-{m^{-1/\beta}}$. Both $CV^2$ and ${GI}_{m}$ are then uniquely determined by the shape parameter which provides a simple handle on the dispersion of the Weibull distribution---the larger $\beta$ the smaller $CV^2$ and ${GI}_{m}$. The criterion in Eq. (\ref{main}) can then be explored graphically (\fref{fig5}) and also written explicitly to read $m>\big[ 2\Gamma(1+1/\beta)^2/\Gamma(1+2/\beta) \big]^\beta \text{ .}$ This inequality relates the branching index $m$ to the shape parameter $\beta$ and provides a simple test to determine the effect of restart with branching. In particular, note that for $\beta<1$, a.k.a. the stretched exponential regime \cite{SEx1,SEx2}, this condition is always satisfied (\fref{fig5} left). In \cite{SM} we demonstrate that  explicit forms of \eref{main} can also be attained for distributions other than the Weibull, e.g., the uniform and Pareto. 

\textit{Discussion.}---The universal criterion in \eref{main} can be re-written in the form 
\bea
\frac{1}{2}\langle T^2 \rangle/\langle T \rangle > \langle \text{min}\{T_1,\cdots,T_m \} \rangle~. \label{Renewal}
\eea
This form has a probabilistic interpretation which becomes evident when contrasting two possible scenarios. Consider first a process (with no restart) that repeats itself indefinitely a-la Sisyphus. Renewal theory then asserts that if this process is probed at an arbitrary time epoch, $\frac{1}{2} \langle T^2 \rangle/\langle T \rangle$ units of time will pass, on average, before the next completion event is observed \cite{Gallager}. Now consider the introduction of an infinitesimally small restart rate. Properties of the Poisson process then assert that restart will occur at an arbitrary moment in time, thus replacing an expected mean completion time of $\frac{1}{2} \langle T^2 \rangle/\langle T \rangle$ pre-restart by $\langle \text{min}\{T_1,\cdots,T_m \} \rangle$ post-restart. Comparing the two alternatives, \eref{main} follows immediately.  

Equation (\ref{Renewal}) invites a connection to the fundamental theorem of Extreme Value Theory (EVT), which pinpoints the universal limit-laws of the maxima and minima of IID random variables: Gumbel, Fr\'echet, or Weibull. In our case, considering a linear scaling of the minimum appearing on the right-hand side of Eq. (\ref{Renewal}), we arrive at the Weibull limit-law. Hence, for $m\gg1$, the minimum is approximated (in law) by $W/s_m$, where $s_m$ is a scaling parameter that satisfies $\underset{m \to \infty}{\lim} s_m=\infty$, and where $W$ is a Weibull-distributed limiting random variable \cite{SM}. Consequently, the right-hand side of \eref{Renewal} can be approximated by $\langle W \rangle /s_m$ -- thus establishing an asymptotic version of the criterion in \eref{Renewal}.

Equation (\ref{main}) can also be expressed in terms of two equality indices. The first equality index is based on the coefficient of variation, and is given by $\mathcal{R}_{2}=1/(1+CV^{2})$; this index belongs to a continuum of equality indices that are intimately related to the Renyi entropy \cite{RenEq}. The second equality index is the counterpart of the generalized Gini index of \eref{Gini}, and is given by $\mathcal{G}_{m}=1-{GI}_{m}$; this index belongs to a continuum of equality indices that extend the Gini index \cite{Iddo-Physica}. In terms of these two equality indices, \eref{main} admits the following product representation
\bea
\mathcal{R}_{2}\cdot \mathcal{G}_{m}<\frac{1}{2}\text{ .}  
\label{equality}
\eea
Interestingly, \eref{equality} resembles an uncertainty principle which provides a fundamental bound on $\mathcal{G}_{m}$ given $\mathcal{R}_{2}$ and vice versa. Indeed, if the introduction of an infinitesimally small restart rate is observed to expedite the completion of a given process: $\mathcal{G}_{m}$ cannot exceed $\frac{1}{2}\mathcal{R}^{-1}_{2}$. Thus, high equality (less uncertainty) of one measure must be compensated by low equality (more uncertainty) of the other. 

\textit{Conclusions.}---Restart has recently attracted considerable attention \cite{Restart1,Restart2,Restart3,Restart4,Restart5,ReuveniPRL,PalReuveniPRL,Restart-Search1,Restart-Search2,Restart-Search3,Iddo-Branch,restart_conc1,restart_conc2,restart_conc3,restart_conc4,restart_conc5,restart_conc6,restart_conc7,restart_conc8,restart_conc9,restart_conc10,restart_conc11,restart_conc12,restart_conc13,restart_conc14,restart_conc15,restart_conc16,restart_conc17,restart_conc18,restart_conc19,restart_conc20,restart_conc21} touching on subjects ranging from stochastic thermodynamics \cite{restart_thermo1,restart_thermo2} to optimization theory \cite{Optimization}, and from quantum mechanics \cite{Quantum1,Quantum2} to biological physics \cite{Restart-Biophysics1,Restart-Biophysics2,Restart-Biophysics3,Restart-Biophysics4}. 
In this letter we extended the theory of first passage under restart to include branching, and provided a universal criterion to determine the effect that restart with branching has on a general FPT process. Throughout this letter it was assumed that restart is global, i.e., that there is one central timer that sets the restart-\&-branching epochs for all the active processes. However, situations where restart is local, i.e., each FPT process has its own internal timer for restart-\&-branching (Fig. S3), could also arise and are as important.

Consider, for example, the spread of an epidemic within a large population of agents (humans, computers, etc.). There, one may be interested in the time it takes for a specific target agent to become infected. The epidemic initiates from an arbitrary agent, `patient zero'. Then, two scenarios can take place: patient zero either infects the target agent before it infects some other agent, or vice versa. The first scenario ends the process. The second scenario generates two infected agents, each proceeding independently the `hunt' for the target agent. Thus, we obtain first passage under local restart-\&-branching (with $m=2$). 

Evidently, global and local  restart-\&-branching are inherently different. Nonetheless, we show that Eqs. (5-7) are not sensitive to this difference \cite{SM}. Namely, the novel universal criterion established in this letter --- to determine the effect of restart-\&-branching on mean completion times --- applies regardless of whether restart is global or local. 

\textit{Acknowledgments.}---We thank Dr. Somrita Ray and Dr. Debasish Mondal for commenting on early versions of this manuscript. A.P. acknowledges support from the Raymond and Beverly Sackler Post-Doctoral Scholarship. S.R. acknowledges support from the Azrieli Foundation.

\widetext

\newpage

\pagebreak
\begin{widetext}

\setcounter{equation}{0}
\setcounter{figure}{0}
\setcounter{table}{0}
\setcounter{page}{1}
\setcounter{section}{0}
\makeatletter
\renewcommand{\theequation}{S\arabic{equation}}
\renewcommand{\thefigure}{S\arabic{figure}}
\renewcommand{\thesection}{S\Roman{section}} 
\renewcommand{\bibnumfmt}[1]{[S#1]}
\renewcommand{\citenumfont}[1]{S#1}



\begin{center}\Large{Supplementary Material for First Passage Under Restart with Branching}\end{center}


\section{Supplementary figures}

\begin{figure}[h]
\centering
\includegraphics[width=8cm]{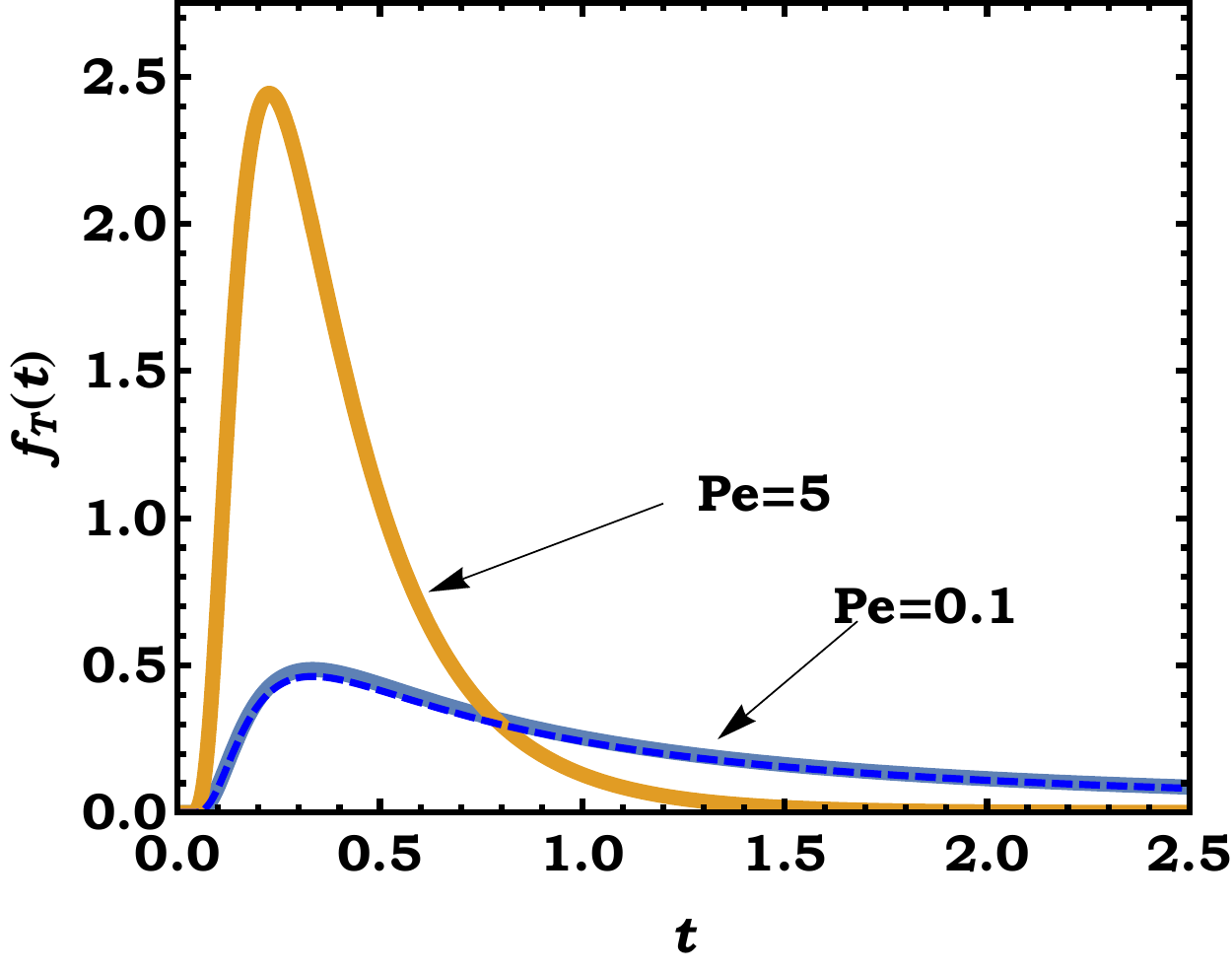}
\caption{The Inverse-Gaussian first passage time distribution for diffusion with drift is plotted for two different values of the P\'eclet number (solid lines). Here, we fixed $L/D=2.0$ so that $Pe=V$. It can be seen that when $Pe=5$ the corresponding distribution is already sharply peaked around its mode. On the other extreme, when $Pe=0.1$, the corresponding distribution closely resembles the one which is obtained in the pure diffusion limit (dashed line).}
\label{figS1}
\end{figure}



\begin{figure}[h]
\centering
\includegraphics[width=8cm]{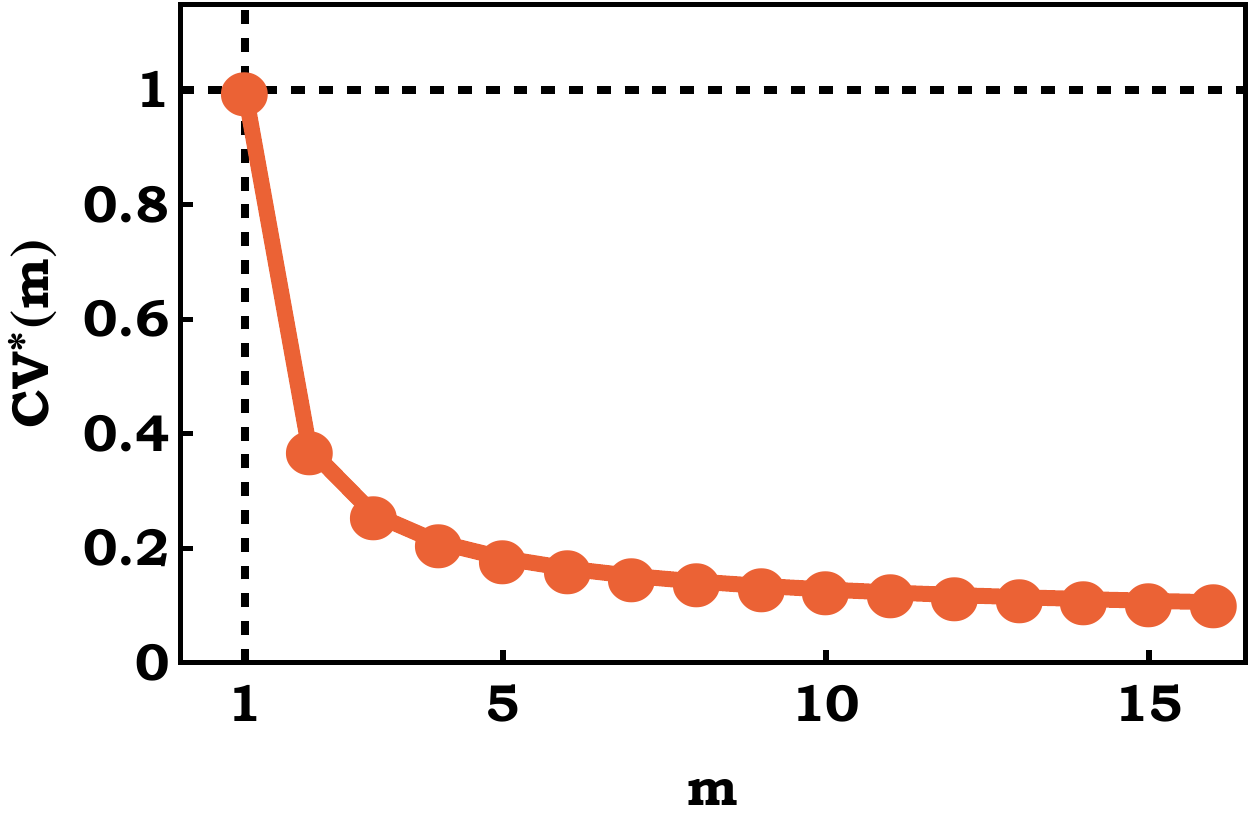}
\caption{Critical $CV$s, as obtained in \fref{fig2} from the relation $CV^2=1/Pe$, vs. the branching index $m$ for diffusion with drift under restart with branching. As $m$ increases, lower $CV$s are found indicating that restart with branching can expedite completion even if first passage time distributions are narrow ($CV<1$). However, while the critical $CV$ value found for $m=1$ is universal (true for any process), the ones found for $m>1$ are particular to diffusion with drift.}
\label{figS2}
\end{figure}

\newpage

\section{Supplementary text}

\subsection{Details of the simulations in \fref{fig2} and \fref{fig4}}

In this subsection, we provide details of simulations done in \fref{fig2} and \fref{fig4}. These were carried out using the following methodology. Based on the process at hand (see details below), we determined the distribution of the random completion time $T$ in the absence of restart and branching. Then, we set the distribution of the restart time $R$ to be exponential with mean $1/r$ (note, again, that this is equivalent to setting a constant restart rate $r$). After these distributions have been set, we simply followed  \eref{renewal-1} to gather statistics about the  completion time of the process  when it is subjected to restart with branching. This was done using the following algorithm: (i) draw two random numbers from the respective distributions of $T$ and $R$; (ii) If $T<R$, set the completion time $T_R$ to $T$ and end the program; (iii) Otherwise, when $R \leq T$, record the time $R$ that has passed, restart the parent process and branch it into $m$ daughter processes that are identical to it; (iv) Set $T_{R}=R+[T^{(m)}]_{R'}$, where $T^{(m)}$ is a random variable distributed like the minimum of $m$ copies of $T$, and $R'$ is a random restart time taken from the same distribution as $R$; (v) Repeat steps (i)-(iv) with $T^{(m)}$ substituting for $T$ to determine $[T^{(m)}]_{R'}$ and consecutively $T_R$. (vi) Repeat steps (i)-(v) many times to gather multiple samples of the random variable $T_R$. Use these to compute $\langle T_R \rangle$---the mean completion time of the process when subjected to restart with branching. Actual implementation of the above algorithm depends on the exact distribution of the random variable $T$ and on the value of the branching index $m$ used. These are specified below.  \\

\begin{itemize}

\item \textbf{\fref{fig2}} --- To study the effect that the introduction of an infinitesimally small restart rate has on the mean FPT of a particle undergoing diffusion with drift, we simulated the process illustrated in \fref{fig1} in the main text. For this, we let the random time $T$ come from the inverse Gaussian distribution \cite{sm-Cox-Miller-Book}
\begin{equation}
f_T(t)=\frac{L}{\sqrt{4\pi D t^3}} e^{-\frac{(L-Vt)^2}{4Dt}}~(t\geq0).
\end{equation} Parameter values were set to: $D=0.5$ and $L=1.0$; and the value of $V$ was set to obey $Pe=LV/2D$. The value of the branching index was varied in the range $m={1,2,3,4}$.  

\item \textbf{\fref{fig4}: Cover time of a random walker} --- Here, we let the random time $T$ be the normalized cover time of a random walker on a network of $N$ sites. The cover time $\mathcal{T}(N)$ is defined as the time needed by the walker to visit all sites in the network. In \cite{sm-Benichou-covertime}, the authors showed that after  proper scaling this cover time admits a universal limit law. More specifically, letting $\langle \tau \rangle$ denote the global mean first passage time, defined as the mean first-passage time to a given target site averaged over all starting sites, it was shown that $X=\mathcal{T}/\langle \tau \rangle-\ln N$ 
converges to the following asymptotic distribution
\bea
f_X(x)=\exp \left( -x -e^{-x} \right)~(-\infty\leq x \leq\infty),
\label{S2}
\eea
in the limit of $N \to \infty$. Setting  \bea
T=\langle\tau\rangle\cdot[X+\ln N]/\langle\mathcal{T}\rangle=\mathcal{T}/\langle\mathcal{T}\rangle,
\label{S3}
\eea
with $N=10^4$ and $\langle\tau\rangle=4.1$, we generated samples from the random variable $T$ by drawing a random number from the distribution of $X$ in \eref{S2} and then substituting it back in \eref{S3}.\\

\item \textbf{\fref{fig4}: Diffusion with drift} --- Here we once again let the random time $T$ come from the inverse Gaussian distribution \cite{sm-Cox-Miller-Book}
\begin{equation}
f_T(t)=\frac{L}{\sqrt{4\pi D t^3}} e^{-\frac{(L-Vt)^2}{4Dt}}~(t\geq0).
\end{equation}
Parameters values were taken  such that $L^2/2D=10.0$ and $L/V=1.0$ thus setting $\langle T \rangle=1.$
\\
\item \textbf{\fref{fig4}: Cell growth model} ---
Here we let the random time $T$ come from the Gamma distribution \bea
f_T(t)=\frac{\kappa e^{-\kappa t}(\kappa t)^{-1+\theta}}{\Gamma(\theta)}~ (t\geq0).
\eea
Parameter values were set to $\kappa=\theta=1/9$ so that $\langle T \rangle=\theta/\kappa=1$. In \cite{sm-Iyer-Biswas} this distribution was used to describe the time it takes a cell growing at a rate $\kappa$ to reach a certain threshold size $\theta$.
\end{itemize}

\subsection{Detailed derivation of \eref{MFPTUB}}
Here we provide a detailed derivation of \eref{MFPTUB} in the main text. We start by letting ${R',R'',R'''...}$ denote independent and identically distributed copies of the random restart time $R$. Iterative use of \eref{renewal-1} in the main text then gives
\bea
T_R &=& \text{min}(T, R)+I(R \leq T)[T^{(m)}]_{R'}, \nn \\
{[T^{(m)}]_{R'}} &=& \text{min}(T^{(m)}, R')+I(R' \leq T^{(m)})[T^{(m^2)}]_{R''}, \nn \\
{[T^{(m^2)}]_{R''}} &=& \text{min}(T^{(m^2)}, R'')+I(R'' \leq T^{(m^2)})[T^{(m^3)}]_{R'''}, \nn \\
&& \hspace{1.5cm}   . \nn \\
&& \hspace{1.5cm}   . \nn \\
&& \hspace{1.5cm}   .
\label{SI1}
\eea
Taking expectations in \eref{SI1} and accommodating all terms in the series, we obtain
\bea
\langle T_R \rangle = \langle \text{min}(T, R) \rangle+\text{Pr}(R \leq T)~\langle \text{min}(T^{(m)}, R) \rangle+
\text{Pr}(R \leq T)~\text{Pr}(R \leq T^{(m)})~\langle \text{min}(T^{(m^2)}, R) \rangle + \cdots~,
\label{SI2}
\eea
where we have used statistical independence, the fact that ${R',R'',R'''...}$ are independent and identically distributed copies of $R$, and the fact that $\langle I(\text{Event}) \rangle=\text{Pr(Event occurred)}$. It will be convenient to define
\bea
p_k&=&\text{Pr}(R\leq T^{(m^k)})~, \nn \\
q_k&=&\langle  \text{min}(T^{(m^k)},R) \rangle.
\label{pq}
\eea
Using the above definitions, we can rewrite the expression for mean completion time from \eref{SI2} as
\bea
\langle T_R \rangle =q_0+~\sum_{n=0}^{\infty}~q_{n+1}\prod_{k=0}^{n}~p_k~.
\label{SI2-1}
\eea
In particular, note that when $R$ is an exponentially distributed random variable with rate $r$, further simplifications can be made by making use of a relation between $p_k$ and $q_k$. In the following, we first derive this relation, and then show how \eref{MFPTUB} is obtained from \eref{SI2-1}. First we observe that
\bea
p_k&=&\text{Pr}(R \leq T^{(m^k)}) \nn \\
&=&  \int_0^\infty ~dt~f_R(t) \text{Pr}(t \leq T^{(m^k)}) \nn \\
&=& r\int_0^\infty ~dt~ \text{Pr}(t \leq T^{(m^k)})~e^{-rt}~,
\label{SI3}
\eea
where we have used the fact that the restart time distribution is exponential i.e., $f_R(t)=re^{-rt}$.
Making use of the same property, we find
$\langle \text{min}(T, R) \rangle =\int_0^\infty~dt~\text{Pr}(\text{min}(T, R) \geq t) = \int_0^\infty~dt~\text{Pr}(T \geq t)~\text{Pr}(R>t)=\int_0^\infty~dt~\text{Pr}(T \geq t)~e^{-rt}$. 
Similarly, one can write
\bea
q_k &=& \langle \text{min}(T^{(m^k)}, R) \rangle  \nn \\
&=& \int_0^\infty~dt~\text{Pr}(T^{(m^k)} \geq t)~e^{-rt}~.
\label{SI4}
\eea
Examining \eref{SI3} and \eref{SI4}, we establish the following relation $q_k=\frac{1}{r}~p_k$. Using the above relation in \eref{SI2-1}, we obtain the following expression for the mean completion time under restart with branching at a constant rate $r$
\bea
\langle T_R \rangle &=&\f{1}{r}~p_0+\f{1}{r}~p_0 p_1+\f{1}{r}~p_0 p_1 p_2+\cdots \nn \\
&=& ~\f{1}{r}~\sum_{n=0}^{\infty}~\prod_{k=0}^{n}~p_k~.
\label{SI6}
\eea
Now using the definition of $p_k$ from \eref{pq}, we can write 
\bea
p_k&=&~\text{Pr}(R \leq T^{(m^k)}) \nn \\ 
~&=&~ r~\int_0^\infty ~dt~ \text{Pr}(t \leq T^{(m^k)})~e^{-rt} \nn \\
~&=&~ 1-\tilde{T}^{(m^k)}(r) ~,
\label{SI6'}
\eea
where $\tilde{T}^{(m^k)}(r)$ is the Laplace transform of $T^{(m^k)}$ evaluated at $r$. 
To arrive at the final expression in \eref{SI6'}, we have used integration by parts in the second line. Substituting \eref{SI6'} in \eref{SI6}, we arrive at \eref{MFPTUB} in the main text
\bea
\langle T_R \rangle &=& \frac{1}{r}~\sum_{n=0}^{\infty}~\prod_{k=0}^{n}~\left[ 1-\tilde{T}^{(m^k)}(r) \right]~.
\label{SI6''}
\eea
In particular, for $m=1$, one can sum the infinite series in \eref{SI6''} to show that it boils down to the result  derived in \cite{sm-ReuveniPRL,sm-PalReuveniPRL}
\bea
\langle T_R \rangle~= ~ \frac{1}{r}~\frac{1-\tilde{T}(r)}{\tilde{T}(r)}~,
\eea
where $\tilde{T}(r)$ is the Laplace transform of $T$ evaluated at $r$.


\subsection{Detailed derivation of \eref{main} in the main text}  
Here we provide a detailed derivation of \eref{main} in the main text. We start with the case of $m=2$. Substituting the definition of $p_k$ in \eref{SI3} into \eref{SI6} we obtain
\bea
\langle T_R \rangle &=& \int_0^\infty~dt~\text{Pr}(T \geq t)~e^{-rt}+ 
r~\int_0^\infty~dt~\text{Pr}(T \geq t)~e^{-rt}~\int_0^\infty~dt~\text{Pr}(T^{(2)} \geq t)~e^{-rt} \nn \\
&+&
r^2~\int_0^\infty~dt~\text{Pr}(T \geq t)~e^{-rt}~\int_0^\infty~dt~\text{Pr}(T^{(2)} \geq t)~e^{-rt}~\int_0^\infty~dt~\text{Pr}(T^{(4)} \geq t)~e^{-rt}~+\cdots
\label{SI7}
\eea
Differentiating both sides of \eref{SI7}
with respect to $r$, we find that $\f{d\langle T_R \rangle}{dr}|_{r=0} < 0$ if and only if
\bea
\int_0^\infty~dt~t~\text{Pr}(T \geq t) > \int_0^\infty~dt~\text{Pr}(T \geq t) \int_0^\infty~dt~\text{Pr}(T^{(2)} \geq t)~.
\label{SI8}
\eea
We now observe that the first term on the right hand side of \eref{SI8} is nothing but $\langle T \rangle$. Indeed,
\bea
\langle T \rangle &=& \int_0^\infty~dt~t~f_T(t)
\nn \\
&=& - \int_0^\infty~dt~t~\f{d}{dt}~\text{Pr}(T \geq t)~
\nn \\
&=&\int_0^\infty~dt~\text{Pr}(T \geq t)~,
\label{SI8-1}
\eea
where to arrive at the last line we performed integration by parts in the second line. 
Similarly, we see that the second term on the right hand side of \eref{SI8} is given by 
\bea
\int_0^\infty~dt~\text{Pr}(T^{(2)} \geq  t)=\int_0^\infty~dt~\text{Pr}(\text{min}\{T_1,T_2\} \geq  t)=\langle \text{min}\{T_1,T_2\}\rangle~.
\label{SI8''}
\eea
As for the term on left hand side of \eref{SI8}, we first note that 
\bea
\langle T^2 \rangle &=& \int_0^\infty~dt~t^2~f_T(t) \nn \\
&=& - \int_0^\infty~dt~t^2~\f{d}{dt}~\text{Pr}(T \geq t)~,
\eea
and integration by parts then gives
\bea
\f{1}{2}~\langle T^2 \rangle= \int_0^\infty~dt~t~\text{Pr}(T \geq t)~.
\label{SI8'}
\eea
Substituting \eref{SI8-1}, \eref{SI8''}, and \eref{SI8'} into \eref{SI8}, we find that the condition in \eref{SI8} is equivalent to
\bea
\f{1}{2}~\langle T^2 \rangle > ~\langle T \rangle~\langle \text{min}\{T_1,T_2\} \rangle~.
\label{SI9}
\eea
Recalling that $CV^2=\sigma^2(T)/\langle T \rangle^2$ and the definition of the Gini index from \eref{Gini} in the main text, we find that  \eref{SI9} boils down to \eref{main} in the main text for $m=2$. 

The result above can be simply extended to the case of a general branching index $m$. To see this, use the same procedure as before to obtain 
\bea
\langle T_R \rangle &=& \int_0^\infty~dt~\text{Pr}(T \geq t)~e^{-rt}+ r~\int_0^\infty~dt~\text{Pr}(T \geq t)~e^{-rt}
~\int_0^\infty~dt~\text{Pr}(T^{(m)} \geq t)~e^{-rt} \nn \\
&+& r^2 \int_0^\infty~dt~\text{Pr}(T \geq t)~e^{-rt}~\int_0^\infty~dt~\text{Pr}(T^{(m)} \geq t)~e^{-rt}~
\int_0^\infty~dt~\text{Pr}(T^{(m^2)} \geq t)~e^{-rt}+\cdots~
\label{S20}
\eea
from \eref{SI6} by use of the definition of $p_k$ in \eref{SI3}. Differentiating both sides of \eref{S20}
with respect to $r$, we find that $\f{d\langle T_R \rangle}{dr}|_{r=0} < 0$ if and only if
\bea
\int_0^\infty~dt~t~\text{Pr}(T \geq t) > \int_0^\infty~dt~\text{Pr}(T \geq t)  \int_0^\infty~dt~\text{Pr}(T^{(m)} \geq t)~.
\label{S21}
\eea
As before, we recall that $\langle \text{min}\{ T_1,\cdots,T_m  \} \rangle=\int_0^\infty~dt~\text{Pr}(T^{(m)}>t)$ which after rearrangement gives us the condition reported in \eref{main} of the main text
\bea
CV^2 + 2 \cdot GI_{m} >1~.
\label{S22}
\eea


\subsection{Explicit forms of the criterion in \eref{main} in the main text}

In this section we consider three different distributions of the time $T$ to show how the criterion in \eref{main} can be computed explicitly.

\subsubsection{Case A: Weibull distribution}
Consider the Weibull distribution $\text{Pr}(T \geq t) = e^{-(t/\alpha)^{\beta}}~(t>0)$, and note that in this case we have  
\bea
\text{Pr}(T^{(m)} \geq t) &=& \text{Pr}(\min \left\{ T_{1},\cdots,T_{m}\right\} \geq t) \nonumber \\
&=& \text{Pr}(T \geq t)^m \nonumber \\
&=& e^{-m(t/\alpha)^{\beta}} \nonumber \\
&=& e^{-(t/\tilde{\alpha})^{\beta}}~,~~~\text{where}~~\tilde{\alpha}=\alpha/m^{1/\beta}.
\eea
We thus see that $T^{(m)}$ is also a Weibull random variable with scale parameter $\tilde{\alpha}$ and shape parameter $\beta$. The expectation of $T^{(m)}$ is then given by $\langle T^{(m)} \rangle=\tilde{\alpha}\Gamma(1+1/\beta)$, where
$\Gamma(x)$ is the Gamma function. Plugging this expression in \eref{Gini}, we obtain
\bea
 GI_{m} &=& 1-\frac{\langle T^{(m)} \rangle}{\langle T \rangle} \nonumber \\
 &=& 1-\frac{1}{m^{1/\beta}}~,
\eea
as stated in the main text. Recalling that the coefficient of variation of the Weibull distribution is given by  $CV^2=\Gamma(1+2/\beta)/\Gamma(1+1/\beta)^{2}-1$, we conclude that for this distribution the criterion in \eref{main} reads
\bea
m>\bigg[ \frac{2\Gamma(1+1/\beta)^2}{\Gamma(1+2/\beta)} \bigg]^\beta \text{ .}
\eea

\subsubsection{Case B: Pareto distribution}

Consider the Pareto distribution $\text{Pr}(T \geq t) = \Big(\frac{1}{1+t}\Big)^{\alpha}$ $(t\geq0)$, with $\alpha>0$ standing for the shape parameter. Recall that the expectation of $T$ diverges for $0<\alpha\leq1$, and that its variance diverges for $1<\alpha\leq2$. In both these cases, and as discussed in the main text, the introduction of restart will always reduce the mean completion time and the same can thus also be said for restart with branching. We therefore focus on the more interesting case of $\alpha>2$. For this we observe that
\bea
\text{Pr}(T^{(m))} \geq t) &=& \text{Pr}(\min \left\{ T_{1},\cdots,T_{m}\right\} \geq t) \nonumber \\
&=& \text{Pr}(T \geq t)^m \nonumber \\
&=& \Big(\frac{1}{1+t}\Big)^{\alpha m} \nonumber \\
&=& \Big(\frac{1}{1+t}\Big)^{\tilde{\alpha}},~~~\text{where}~~\tilde{\alpha}=\alpha m.
\eea
From here we see that $T^{(m)}$ is also a Pareto random variable with shape parameter $\tilde{\alpha}$. The expectation of $T^{(m)}$ is then given by $\langle T^{(m)} \rangle = 1/(\tilde{\alpha}-1)$. Substituting this expression into \eref{Gini} in the main text, we find
\bea
 GI_{m} &=& 1-\frac{\langle T^{(m)} \rangle}{\langle T \rangle} \nonumber \\
 &=& \frac{\alpha m - \alpha}{\alpha m-1}~.
\eea
For $\alpha>2$ the coefficient of variation of the Pareto distribution is well defined and given by $CV^2=\alpha/(\alpha-2)$. Equation (\ref{main}) in the main text boils down to
\bea
m~>~\frac{\alpha-1}{\alpha}~.
\eea

\subsubsection{Case C: Uniform distribution}

Consider the uniform distribution on the unit interval $[0,1]$. For this distribution $\text{Pr}(T \geq t) = (1-t)$ ($0\leq t \leq 1$); and we thus have 
\bea
\text{Pr}(T^{(m)} \geq t) &=& \text{Pr}(\min \left\{ T_{1},\cdots,T_{m}\right\} \geq t) \nonumber \\
&=& \text{Pr}(T \geq t)^m \nonumber \\
&=& (1-t)^m~,~0\leq t \leq 1.
\eea
It is then easy to show that $\langle T^{(m)} \rangle= \frac{1}{m+1}$ and  $GI_{m}=\frac{m-1}{m+1}$ follows immediately. In addition, the coefficient of variation for this distribution is simply given by $CV^2=1/3$. Equation (\ref{main}) in the main text then boils down to a particularly simple condition \bea
m~>~2~.
\eea


\subsection{Connection with extreme value theory}

In this section we derive an asymptotic version of \eref{Renewal} by use of extreme value theory. Denoting the cumulative distribution function of the random variable $T$ by $F\left( t\right)=\text{Pr}(T\leq t)~(t\geq 0)$; we assume that $F(t)$ is regularly varying at the origin, i.e., that \cite{sm-regular-variation}
\bea
\lim_{l\rightarrow 0}\frac{F\left( l t \right) }{F\left( l \right) }=t^{\epsilon }~,  \label{21}
\eea
for some $\epsilon>0$ (the regular-variation
exponent). For example, it is easy to verify that the cumulative distribution functions of the Weibull, Pareto and Uniform distributions in the previous section all have this property. To see this, simply observe that: (i) for the Weibull--- $\underset{l \to 0}{\lim} \frac{F\left( l t \right) }{F\left( l \right) }=t^{\beta }$, so that $\epsilon=\beta>0$; (ii) for the Pareto---$\underset{l \to 0}{\lim} \frac{F\left( l t \right) }{F\left( l \right) }=t$, so that $\epsilon=1$; and (iii) for the Uniform--- $\underset{l \to 0}{\lim} \frac{F\left( l t \right) }{F\left( l \right) }=t$, so that  $\epsilon=1$. 

Examine now the random variable $W_{m} \equiv s_m \cdot \min \left\{ T_{1},\cdots,T_{m}\right\}$ (where $s_m$ is a positive scalar parameter) which satisfies 
\begin{equation}
\lim_{m\rightarrow \infty }m\cdot F\left( \frac{1}{s_m}\right) =1\text{ ,}
\label{22}
\end{equation}%
e.g., take $s_m$ such that $m\simeq 1/F\left( 1/s_m \right)$ for $m \gg 1$. Extreme value theory then asserts that \cite{RT}
\begin{equation}
\lim_{m\rightarrow \infty }\Pr \left( W_{m}>t\right) =\exp \left(
-t^{\epsilon }\right)~(t\geq 0),   \label{23}
\end{equation}%
i.e., $W_{m}$ converges, in law, to a limit governed by the Weibull distribution. One can then approximate $\langle W_{m} \rangle \simeq \Gamma (1+\frac{1}{%
\epsilon })$. Recalling \eref{Renewal} in the main text we multiply and divide its right hand side by $s_m$ to obtain 
\begin{equation}
\frac{\langle T^{2} \rangle }{2\langle T \rangle }>\frac{1}{s_m
}\langle  s_m \cdot \min \left\{ T_{1},\cdots ,T_{m}\right\} \rangle=\frac{1}{s_m}\langle W_m \rangle 
\text{ .}  \label{20}
\end{equation}
Substituting the expression for $\langle W_{m} \rangle$ we obtain an asymptotic version of \eref{20}
\begin{equation}
\frac{\langle T^{2} \rangle }{2\langle T \rangle }
>\frac{1}{s_m%
}\Gamma \left( 1+\frac{1}{\epsilon }\right) \text{ .}  \label{24}
\end{equation}%
As the left-hand side of \eref{24} is fixed, and  $\underset{m \to \infty}{\lim} s_m=\infty$ we conclude that, under the conditions specified above, the criterion in \eref{24} can always be met by taking $m$ to be large enough.


\subsection{Proof that Eqs. (5-7) in main text continue to hold when processes restart-\&-branch independently}

Restart and branching can occur in two different ways: global and local. In the former case, there is a central clock that “announces”, for all running FPT processes, the restart-\&-branching epochs. While in the latter case each FPT process has its own internal clock which determines when to restart-\&-branch. The difference between the two cases is schematically illustrated in the figure below. 

\begin{figure}[h]
\centering
\includegraphics[width=8cm]{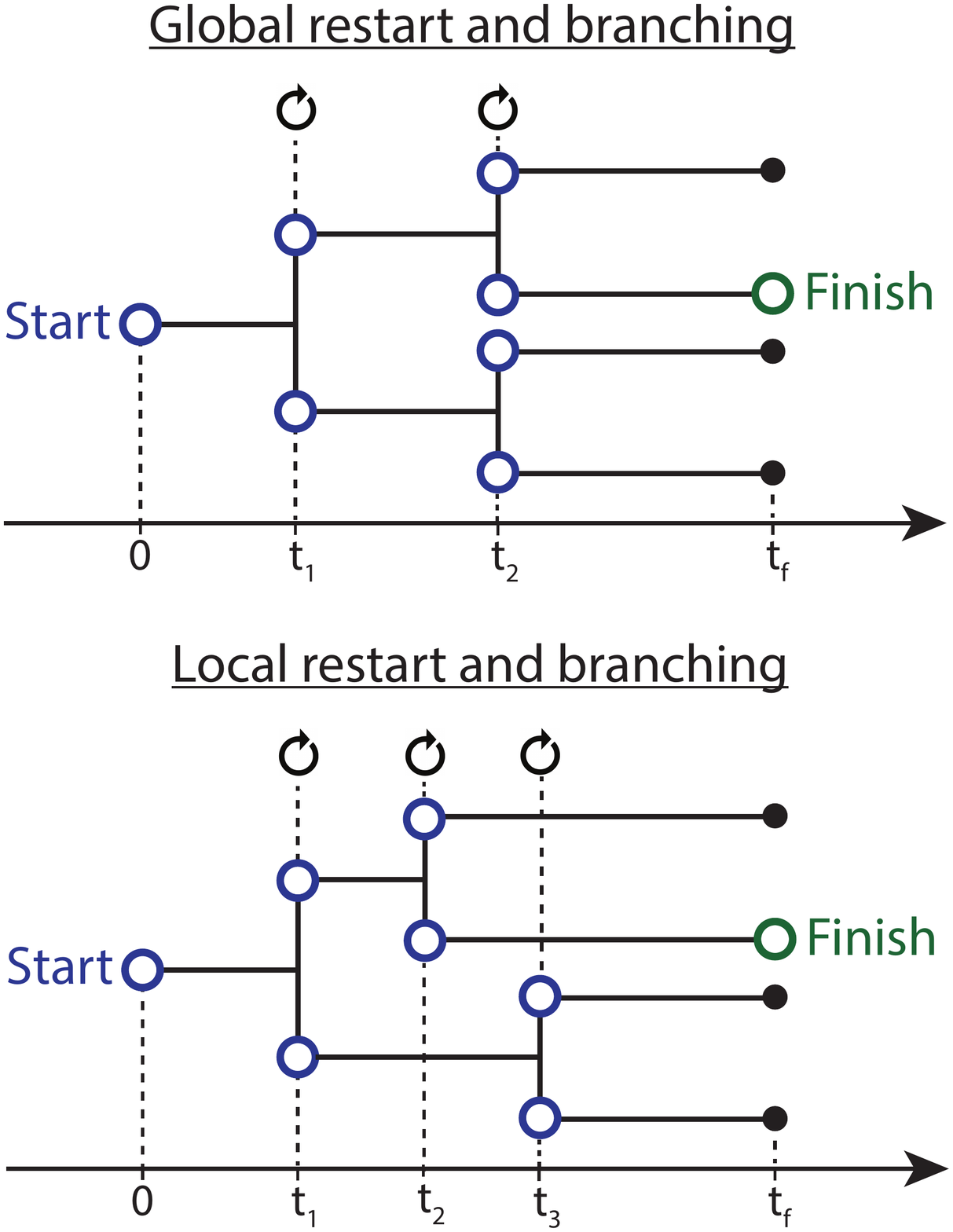}
\caption{Global vs. local restart and branching.}
\label{figS3}
\end{figure}

In the main text, we focused on the case of global restart and branching. However, the results in Eqs. (5-7) hold equally for local restart and branching. To understand this one only needs to recall that the universal criterion in Eq. (5) tells us when will the introduction of restart and branching reduce the mean FPT of a process. To decide on this issue, one needs to determine what happens when an infinitesimally small restart rate is “turned on”. In other words, one needs to consider the limit in which the restart rate goes to zero. However, in this limit restart happens very infrequently, thus rendering global and local restart identical to first order. Indeed, differences between the two cases become discernible from the second restart event onward. But when the restart rate is infinitesimally small it is enough to consider situations where restart did not occur at all or occurred only once. All other events could be safely neglected. To finish, we give a formal proof of this statement.

In the case of local restart and branching one can write a stochastic renewal equation that is analogous to Eq. (1) in the main text. This reads 
\begin{equation}
\begin{array}{l}
T_{R}=\left\{ \begin{array}{lll}
T &  & \text{if }T<R\text{ }\\
 & \text{ \ \ }\\
R+\text{min}\{T_R^1,\cdots,T_R^m\} &  & \text{if }R\leq T\text{ ,}
\end{array}\right.\text{ }\end{array}
\end{equation}  
where $\{T_R^1,\cdots,T_R^m\}$ are independent and identically distributed copies of $T_R$. Taking  expectations on both sides, we obtain
\bea
\langle  T_R \rangle=\langle \text{min}(T,R) \rangle+\text{Pr}(R \leq T) \langle \text{min}\{T_R^1,\cdots,T_R^m\} \rangle \text{ .}
\label{S41}
\eea
To finish, we recall that when the restart time $R$ is exponential with rate $r$ we have $\langle \text{min}(T,R) \rangle=\frac{1-\tilde{T}(r)}{r}$, and
$\text{Pr}(R \leq T)=1-\tilde{T}(r)$. Applying a small $r$ expansion on the right hand side of \eref{S41}, we find
\bea
\langle  T_R \rangle=\langle  T \rangle+r \left[ \langle  T \rangle \cdot \langle  T^{(m)} \rangle -\frac{\langle  T^2 \rangle}{2} \right]
+ \mathcal{O}(r^2)\text{ ,}
\eea
where we recalled that $T^{(m)}=\text{min}\{T_1,\cdots,T_m\}$ and further used the fact that $\langle \text{min}\{T_R^1,\cdots,T_R^m\} \rangle=\langle T^{(m)} \rangle + \mathcal{O}(r)$. Thus, the introduction of restart will expedite the underlying process only when 
\bea
\langle  T \rangle \cdot \langle  T^{(m)} \rangle -\frac{\langle  T^2 \rangle}{2} <0
\eea
which is equivalent to the criterion given by Eq. (5) in the main text.



\end{widetext}

\end{document}